\documentclass[12pt]{iopart}

\usepackage[figuresright]{rotating}
\usepackage{epsfig}
\usepackage{graphicx}
\usepackage{bm}

\newcommand{\AmS}{{\protect\the\textfont2
  A\kern-.1667em\lower.5ex\hbox{M}\kern-.125emS}}
\newcommand{ \srts }{$\sqrt{s_{_{\rm NN}}}$}

\def\Journal#1#2#3#4{{#1} {\bf #2}, #3 (#4)}

\def\NIMA{{ Nucl. Instrum. Methods} A}

\def\PLB{{ Phys. Lett.}  B}
\def\PRL{ Phys. Rev. Lett.}

\def\PRC{{ Phys. Rev.} C}
\def\PRD{{ Phys. Rev.} D}

\def\JPG{{ J. Phys.} G}

\def\EPJC{{ Eur. Phys. J.} C}

\begin{document}

\title[]{Open charm hadron measurement in $p$+$p$ and Au+Au collisions at \srts\ = 200 GeV in STAR}

\author{Yifei Zhang (for the STAR Collaboration)}
\address{Lawrence Berkeley National Lab, Berkeley, CA 94720, USA}
\ead{yzhang6@lbl.gov}

\begin{abstract}

We present the measurements of $D^0$ and $D^*$
in $p$+$p$ and $D^0$ in Au+Au collisions via hadronic decays $D^{0}\rightarrow K^{-}\pi^{+}$, $D^{*+}\rightarrow D^{0}\pi^{+}\rightarrow  K^{-}\pi^{+}\pi^{+}$ in mid-rapidity $|y|<1$ at \srts\ = 200 GeV, covering $p_T$ from 0.2 to 6 GeV/$c$ in $p$+$p$ and 0.4 to 5 GeV/$c$ in Au+Au, respectively. The charm pair production cross section per nucleon-nucleon collision at mid rapidity is measured to be 202 $\pm$ 56 (stat.) $\pm$ 40 (sys.) $\pm$ 20 (norm.) $\mu$b in $p$+$p$ and 186 $\pm$ 22 (stat.) $\pm$ 30 (sys.) $\pm$ 18 (norm.) $\mu$b in Au+Au minimum bias collisions. The number of binary collisions scaling of charm cross section indicates that charm is produced via initial hard scatterings. No suppression of $D^0$ $R_{AA}$ in Au+Au 0-80\% minbias collisions is observed at $p_T$ below 3 GeV/$c$. Blast-wave predictions with light-quark hadron parameters are different from data, which may indicate that $D^0$ decouples earlier from the medium than the light-quark hadrons.

\end{abstract}
\vspace{-0.5cm}

\vspace{-0.2cm}
\section{Introduction}
\vspace{-0.25cm}

Heavy quarks are believed to be an ideal probe to study the properties
of the QCD medium produced in relativistic heavy ion collisions.
Heavy quark production in elementary collisions is expected
to be well calculated in perturbative QCD~\cite{pQCD,lin,FONLL}. Precise
understanding of both the charm production total cross section and the
fragmentation in p+p collisions serves as a baseline for the investigation of 
the QCD medium via open charm and charmonium in heavy ion collisions. Comparing charm differential cross sections per nucleon-nucleon collision between Au+Au and $p$+$p$ collisions can help in understand medium modifications due to interactions with the hot and dense system. It is also of interest to study the D-meson freeze-out and flow properties, which may tell us how the heavy-quark hadrons behave compared to the light-quark hadrons when they decouple from the medium.
Early RHIC measurements were carried out mostly via electrons from heavy quark semi-leptonic decays~\cite{STAREMCe,PHENIXe}. There are limitations in the electron approach: the charm hadron and electron kinematics are only weakly
correlated due to the decay, and measured electrons have mixed
contributions from various charm and bottom hadrons. Thus direct measurement
of charm hadrons via hadronic decays is crucial for a better understanding of 
charm-medium interactions at RHIC. 

In this paper, we present the measurement of open charm hadrons, $D^0$ and $D^*$ in $p$+$p$ and $D^0$ in Au+Au collisions via hadronic decays $D^{0}\rightarrow K^{-}\pi^{+}$ ($BR$ = 3.89\%), $D^{*+}\rightarrow D^{0}\pi^{+}$ ($BR$ = 67.7\%) $\rightarrow  K^{-}\pi^{+}\pi^{+}$ and their charge conjugates at mid-rapidity $|y|<1$ at \srts\ = 200 GeV in the STAR experiment. 

\vspace{-0.1cm}
\section{Analysis}
\begin{figure}[htp]
\vspace{-0.3cm}
\centering
\includegraphics[width=1.8in]{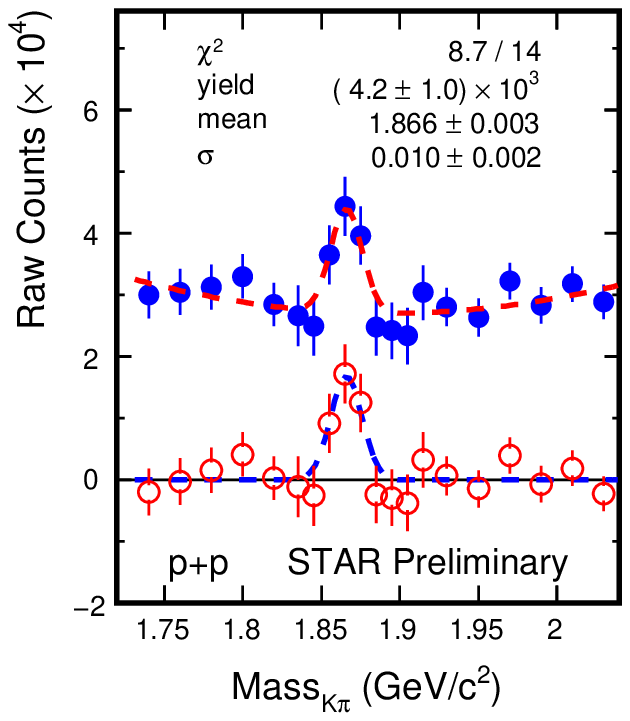}
\includegraphics[width=1.8in]{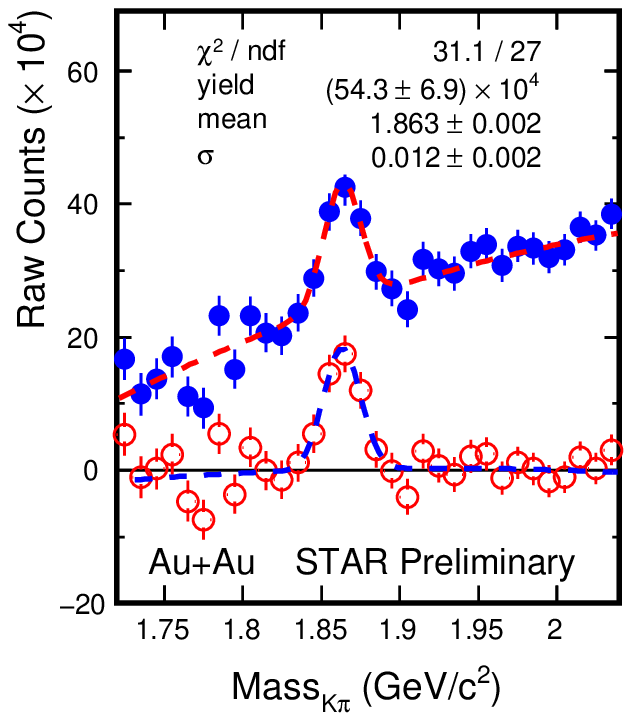}
\includegraphics[width=2.4in]{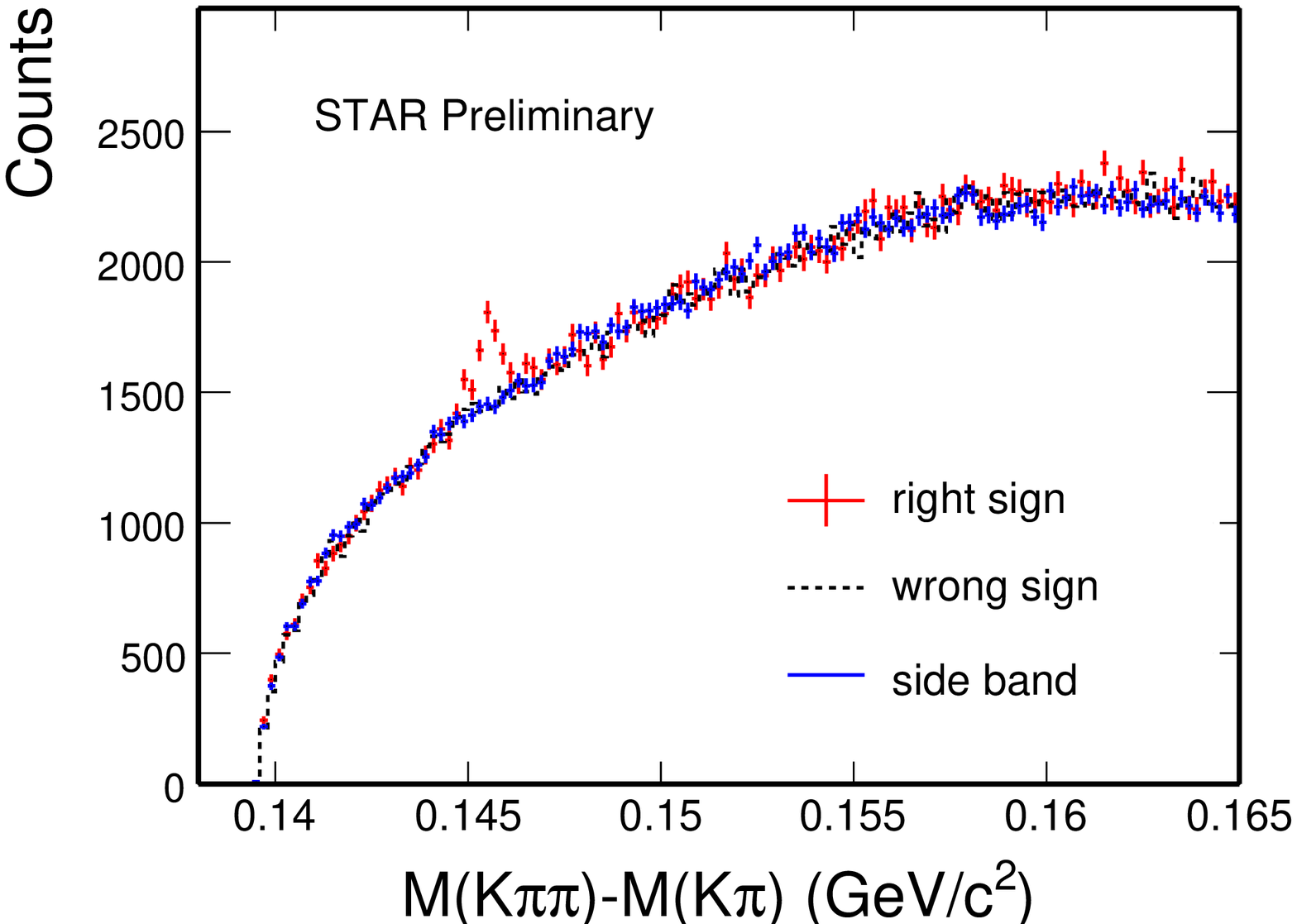}
\vspace{-0.25cm}
\centering \caption{Left and middle panels: $D^0$ signals in $p$+$p$ and Au+Au 0-80\% minbias collisions after same-sign and mix-event background subtraction, respectively. Right panel: $D^*$ signal in $p$+$p$ collisions. Combinatorial background is reproduced by the distributions from the wrong-sign (black dotted) and side-band (blue solid) methods.}\vspace{-0.25cm}\label{fig:Dsignals}
\end{figure}

\vspace{-0.25cm}
The analysis is based on 105 million p+p minimum bias (minbias) events collected at STAR in the year 2009 and 280 million Au+Au minbias events from year 2010. In this analysis we used the time projection chamber (TPC), which provides the information of particle momentum and ionization energy loss ($dE/dx$) covering a large acceptance with full 2$\pi$ azimuthal angle at $|\eta|<1$, and the Time-Of-Flight detector (TOF), which covered 72\% in 2009 and 100\% in 2010 of the whole barrel. The decay daughter (K$\pi$) identification was greatly improved by a combination of TPC $dE/dx$ and TOF measured particle velocity $\beta$~\cite{STARPID}. The $D^0$ was reconstructed via K$\pi$ invariant mass. The same-sign and mix-event background subtraction in $p$+$p$ and Au+Au collisions respectively followed the same analysis techniques as previous analysis~\cite{dAuCharm}. In Fig.~\ref{fig:Dsignals} $D^0$ signals are shown as open circles in both $p$+$p$ ($0.2 < p_{T} < 2.2$ GeV/$c$) and Au+Au 0-80\% minbias ($0.4 < p_{T} < 5.0$ GeV/$c$) collisions. The signal and the residual background shown as the filled circles are described by a Gaussian plus 2nd order polynomial function. $D^*$ ($2.0 < p_{T} < 6.0$ GeV/$c$) was reconstructed via a mass difference between $D^{0}\pi$ and $D^{0}$ in $p$+$p$ collisions based on the analysis techniques in Ref~\cite{starDstar}. The combinatorial background is reproduced by the distributions from the  wrong-sign and side-band methods.

\vspace{-0.3cm}
\section{Results}
\vspace{-0.25cm}

\begin{figure}[htp]
\vspace{-0.4cm}
\centering
\includegraphics[width=2.6in]{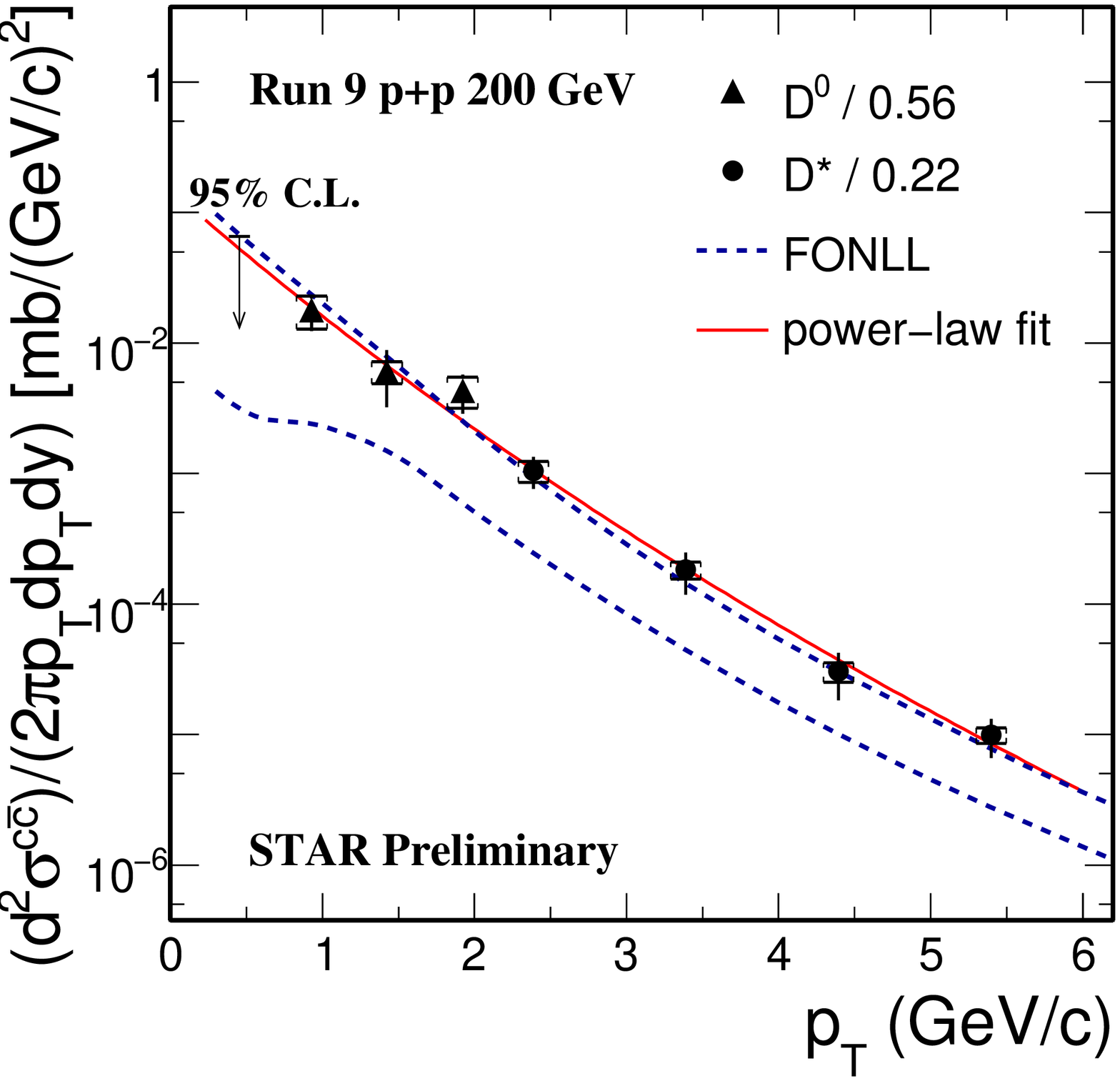}
\includegraphics[width=2.7in]{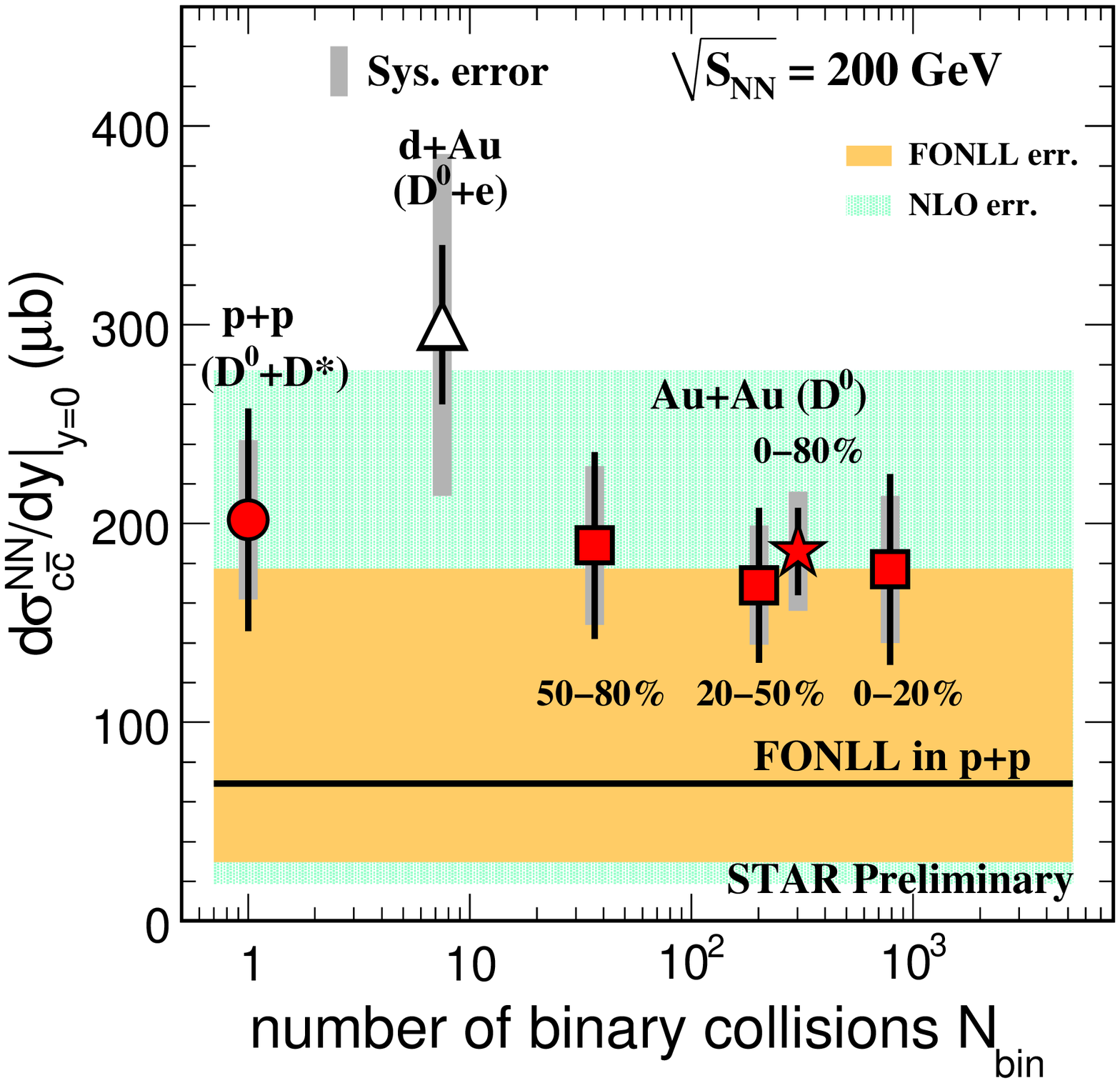}
\vspace{-0.25cm}
\centering \caption{Left Panel: c$\bar{c}$ pair production cross sections (symbols) as a function of $p_T$ in 200 GeV $p$+$p$ collisions. Right Panel: The charm production cross section per nucleon-nucleon collision at mid-rapidity as a function of $N_{bin}$.}\vspace{-0.15cm}\label{fig:ccxsec}
\end{figure}

The full efficiency and acceptance corrected charm pair differential production cross section at mid-rapidity in 200 GeV $p$+$p$ collisions is shown in the left panel of figure~\ref{fig:ccxsec} and compared with Fixed-Order Next-to-Leading Logarithm (FONLL) calculations (dashed curve)~\cite{FONLL}. The Data agree with the FONLL upper bound. $D^0$ (triangles $0.2 < p_{T} < 2.2$ GeV/$c$) and $D^*$ (circles $2.0 < p_{T} < 6.0$ GeV/$c$) are scaled by the charm fragmentation ratios 
~\cite{pdg}. The charm production cross section at mid-rapidity in 200 GeV $p$+$p$ collisions was extracted, from a power-law function fit (solid curve), as 202 $\pm$ 56 (stat.) $\pm$ 40 (sys.) $\pm$ 20 (norm.) $\mu$b. The dominant systematic uncertainties are from $D^0$ background subtraction and the raw yield extraction.

\begin{figure}[htp]
\vspace{-0.3cm}
\centering
\includegraphics[width=4.8in]{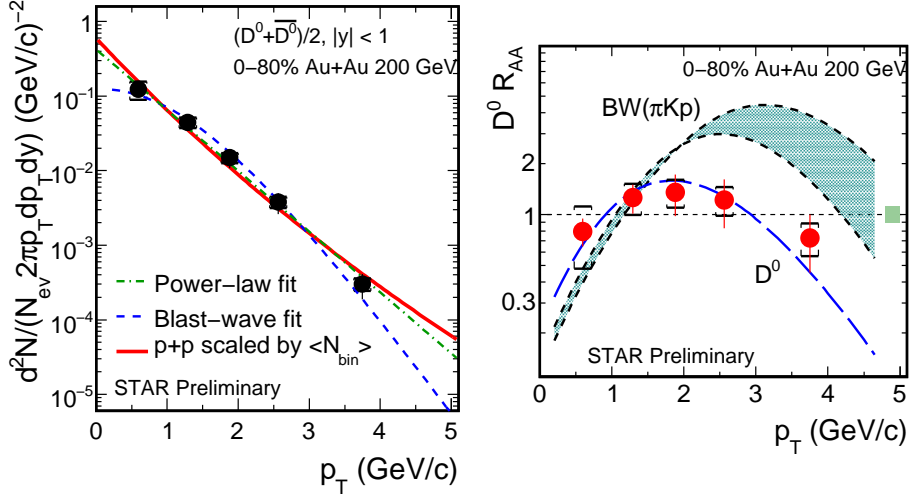}
\vspace{-0.35cm}
\centering \caption{Left Panel: $D^0$ $p_{T}$ spectrum in 200 GeV 0-80\% Au+Au collisions. Right Panel: $D^0$ nuclear modification factor $R_{AA}$ as a function of $p_T$.}\vspace{-0.35cm}\label{fig:D0AuAu}
\end{figure}

The same technique~\cite{dAuCharm} was applied to obtain the $D^0$ invariant $p_T$ distributions in 0-80\% minbias Au+Au collisions shown as the left panel of Fig.~\ref{fig:D0AuAu}. The charm production cross section per nucleon-nucleon collision at mid-rapidity in 200 GeV Au+Au collisions was extracted, from the average of a power-law (dot-dashed curve) and a blast-wave (dashed curve) fit, as 186 $\pm$ 22 (stat.) $\pm$ 30 (sys.) $\pm$ 18 (norm.) $\mu$b, assuming that the ratio of $D^0$ to the number of c$\bar{c}$ pair does not change from $p$+$p$ to Au+Au collisions. The power-law fit to $p$+$p$ scaled by the average number of binary collisions ($N_{bin}$) is shown as solid curve. The charm cross section for three centrality bins, 0-20\%, 20-50\% and 50-80\%, is obtained according to the integrated yields. The charm production cross section per nucleon-nucleon collision at mid-rapidity as a function of $N_{bin}$ is shown in the right panel of Fig.~\ref{fig:ccxsec}. The results in $p$+$p$, $d$+Au~\cite{dAuCharm} and Au+Au 0-80\% minbias collisions are shown as a circle, a triangle and a star. Squares show the results from the three centrality bins. Within errors, the results are in agreement and follow the number of binary collisions scaling, which indicates that charm quark is produced via initial hard scatterings at early stage of the collisions at RHIC. The FONLL (orange band) and NLO~\cite{NLO} (light-blue band) 1-$\sigma$ uncertainties are also shown here for comparison.

The $D^{0}$ nuclear modification factor $R_{AA}$ was obtained via dividing $D^{0}$ yields in Au+Au 0-80\% minbias collisions by the power-law fit to $p$+$p$ yields scaled by $N_{bin}$, shown in right panel of Fig.~\ref{fig:D0AuAu}. The uncertainty of the $p$+$p$ power-law shape is taken into account as systematic error. No suppression is observed at $p_{T} < 3$ GeV/$c$. The dashed curve shows the blast-wave fit. The shaded band is the predicted $D^{0}$ $R_{AA}$ with blast-wave parameters from light-quark hadrons~\cite{BWLH}, which is different from data. This may indicate that $D^0$ mesons freeze out earlier than light-quark hadrons.

\vspace{-0.3cm}
\section{Summary and outlook}
\vspace{-0.25cm}

Open charm hadrons ($D^0$ and $D^*$) are measured in $p$+$p$ and Au+Au collisions at \srts\ = 200 GeV at STAR.
Charm cross sections per nucleon-nucleon collision at mid-rapidity follow the number of binary collisions scaling, which indicates that charm quarks are produced via initial hard scatterings in the early stage of the collisions at RHIC. The charm pair production cross sections per nucleon-nucleon collision at mid rapidity are measured to be 202 $\pm$ 56 (stat.) $\pm$ 40 (sys.) $\pm$ 20 (norm.) $\mu$b in $p$+$p$ and 186 $\pm$ 22 (stat.) $\pm$ 30 (sys.) $\pm$ 18 (norm.) $\mu$b in Au+Au minimum bias collisions. No suppression of $D^0$ $R_{AA}$ in Au+Au 0-80\% minbias collisions is observed at $p_T$ below 3 GeV/$c$. Blast-wave predictions with light-quark hadron parameters are different from data, which may indicate that $D^0$ decouples earlier from the medium than the light-quark hadrons. In the near future the STAR Heavy Flavor Tracker~\cite{HFT} will provide the necessary resolution to reconstruct secondary vertices of D-mesons, which will increase the precision of charm measurements, in particular $D^0$ $v_2$ and high $p_T$ $D^0$ $R_{AA}$, to address the light flavor thermalization and charm quark energy loss mechanisms.

\end{document}